\documentclass[fleqn,usenatbib]{mnras}

\usepackage{newtxtext,newtxmath}
\usepackage{amsmath}

\usepackage[T1]{fontenc}

\DeclareRobustCommand{\JONG}[3]{#2}
\let\JONGthebibliography\thebibliography
\def\thebibliography{\DeclareRobustCommand{\JONG}[3]{##3}\JONGthebibliography}
\DeclareRobustCommand{\MAATEN}[3]{#2}
\let\MAATENthebibliography\thebibliography
\def\thebibliography{\DeclareRobustCommand{\MAATEN}[3]{##3}\MAATENthebibliography}

\usepackage{graphicx}	% Including figure files
\usepackage{amsmath}	% Advanced maths commands
\usepackage{amssymb}	% Extra maths symbols

\title[Follow-up of GALAH DR3 EMP candidates]{Spectroscopic follow-up of statistically selected extremely metal-poor star candidates from GALAH DR3}

\author[G. S. Da~Costa et al.]{G. S. Da Costa,$^{1,2}$\thanks{E-mail: gary.dacosta@anu.edu.au}
M. S. Bessell,$^{1}$
Thomas Nordlander,$^{1,2}$
Arvind C. N. Hughes,$^{2,3,4,5}$
Sven Buder,$^{1,2}$
\newauthor A. D. Mackey,$^{1,2}$, Lee R. Spitler,$^{2,3,4,6}$
and D. B. Zucker$^{2,3,4}$
\\
$^{1}$Research School of Astronomy and Astrophysics, Australian National University, Canberra, ACT 0200, Australia\\
$^{2}$ARC Centre of Excellence for Astrophysics in Three Dimensions (ASTRO-3D), Canberra, ACT 2611, Australia\\
$^{3}$School of Mathematical and Physical Sciences, Macquarie University, Sydney, NSW 2019, Australia\\
$^{4}$Research Centre in Astronomy, Astrophysics and Astrophotonics, Macquarie University, Sydney, NSW 2019, Australia\\
$^{5}$Max Planck Institute for Astronomy, Heidelberg, Germany\\
$^{6}$Australian Astronomical Optics, Faculty of Science and Engineering, Macquarie University, Macquarie Part, NSW 2113, Australia
}

\date{Accepted XXX. Received YYY; in original form ZZZ}

\pubyear{2022}

\begin{document}
\label{firstpage}
\pagerange{\pageref{firstpage}--\pageref{lastpage}}
\maketitle

% Abstract of the paper
\begin{abstract}
The advent of large-scale stellar spectroscopic surveys naturally leads to the implementation of machine learning techniques to isolate, for example, small sub-samples of potentially interesting stars from the full data set. A recent example is the application of the t-SNE statistical method to $\sim$600,000 stellar spectra from the GALAH survey in order to identify a sample of candidate extremely metal-poor (EMP, [Fe/H] $\leq$ --3) stars.  We report the outcome of low-resolution spectroscopic follow-up of 83 GALAH EMP candidates that lack any previous metallicity estimates. Overall, the statistical selection is found to be efficient ($\sim$one-third of the candidates have [Fe/H] $\leq$ --2.75) with low contamination ($<$10\% have [Fe/H] $>$ --2), and with a metallicity distribution function that is consistent with previous work.  Five stars are found to have [Fe/H] $\leq$ --3.0, one of which is a main sequence turnoff star.  Two other stars are revealed as likely carbon-enhanced metal-poor (CEMP) stars of type CEMP-$s$, and a known carbon star is re-identified.  The results indicate that the statistical selection approach employed was successful, and therefore it can be applied to forthcoming even larger stellar spectroscopic surveys with the expectation of similar positive outcomes.
\end{abstract}

% Select between one and six entries from the list of approved keywords.
% Don't make up new ones.

%%%%%%%%%%%%%%%%%%%%%%%%%%%%%%%%%%%%%%%%%%%%%%%%%%

%%%%%%%%%%%%%%%%% BODY OF PAPER %%%%%%%%%%%%%%%%%%

\section{Introduction}

%  full version of Table \ref{tab:tab2} has 76 entries and nothing in common with Table 1 of Hughes2022.

The most metal-poor stars in the Milky Way are key objects for understanding star formation and chemical enrichment at the earliest epochs in the Galaxy's evolution \citep[see, e.g., the reviews of][]{Beers2005,Frebel2015}.  This is because the Big Bang produced only hydrogen, helium and trace amounts of light elements such as lithium: the heavier elements result solely from the evolution of the stars that formed from the original pristine gas, the so-called Population III, or first generation, stars.  The supernova explosions of these stars enriched the surrounding gas with their nucleosynthetic products that were then incorporated into the subsequent generations of stars.  No genuine Population~III star, which would necessarily need to be of sub-solar mass to have survived a Hubble time, has yet been discovered.  Nevertheless, the properties of the first generation stars can be investigated through the study of element abundances in second and third generation stars\footnote{Second generation stars are likely the CEMP-no stars that have substantial overabundances in [C/Fe] coupled with very low iron abundances, while third generation stars are likely those also with very low iron abundances but which lack any significant carbon overabundance, i.e., possess approximately solar [C/Fe] values \citep[e.g.,][]{Hansen2016,Yoon2019,Norris2019}.}, stars that are characterized by very low overall abundance compared that of the Sun \citep[e.g.,][]{Nordlander2019,Kielty2021,Welsh2021}.

The high rate of star formation at early times in the Galaxy's formation with the accompanying rapid rise in stellar and gas metallicities means that second and third generation stars are extremely rare.   The search for, and study of, such stars has nevertheless been an active field for decades.  Searches for extremely metal-poor (EMP) stars, which have [Fe/H] $\leq$ --3.0 \citep{Beers2005}, have employed a variety of techniques \citep[see, e.g.,][for examples of large scale surveys for EMP stars]{DaCosta2019}. They fundamentally fall into one of two categories: searches that target candidates via EMP-specific spectroscopic or photometric characteristics, or searches that seek to identify rarely occurring EMP stars in large scale spectroscopic surveys that have no EMP-specific bias.  With the development of highly multiplexed spectrographs and dedicated facilities, large scale spectroscopic datasets that contain $\sim$10$^5$--10$^6$ objects are now becoming available, with examples including the APOGEE \citep{Ahumada2020} and GALAH \citep{Buder2021} surveys.  The enormous number of stars in these and other similar surveys, combined with the extreme rarity of EMP stars, necessarily requires the implementation of machine-learning techniques to isolate the likely tiny number of potential EMP candidates in such extensive samples.  
One such approach is discussed in \citet{Hughes2022} in which the t-SNE \citep[t-distributed stochastic embedding;][]{Maaten2008} method is applied to the 
$\sim$600,000 stellar spectra in Data Release 3 (DR3) of the GALAH survey \citep{Buder2021}.  Other examples of the application of the t-SNE approach to search for metal-poor stars in stellar spectroscopic survey data include \citet{Matijevi2017} and \citet{Anders2018}.  In essence, the t-SNE method reduces high dimensional data to a 2D map where stars with similar characteristics are close together and dissimilar stars are well separated.  Consequently, stars that fall near known EMP stars in the t-SNE map are likely EMP candidates.  

The GALAH data are derived from spectra obtained with the HERMES high resolution spectrograph \citep{Sheinis2015} at the Anglo-Australian Telescope.  In order to accommodate the multi-object capability provided by the 2dF multi-fibre instrument, HERMES spectra are restricted to four fixed wavelength channels (blue: $\lambda$4713--4903\AA, green: $\lambda$5648--5873\AA, red: $\lambda$6478--6737\AA\/ and IR: $\lambda$7585--7887\AA) that are optimum for the studying abundances of up to 30 elements in the approximately solar metallicity dwarfs that dominate the GALAH sample \citep{Buder2021}.  The consequence of this choice, however, is that GALAH spectra for EMP-stars are largely featureless in the HERMES wavelength channels with the Balmer lines H$\beta$ and H$\alpha$ the only strong features.  For this reason \citet{Hughes2022} did not apply the t-SNE approach to the entire available wavelength space.  Instead the analysis was restricted to wavelength regions around H$\alpha$ and H$\beta$ plus the intervals $\lambda$4867--4872\AA\/ and $\lambda$4887--4892\AA\/ that contain the strongest Fe~{\sc i} lines in the HERMES channels. A region ($\lambda$7771.94--7775.39\AA) from the IR-channel that covers the O~{\sc i} triplet was also included to discriminate against hot stars \citep{Hughes2022}.  As discussed in \citet{Hughes2022}, the analysis resulted in $\sim$2500 stars (less than 1\% of the input sample!) in a ``metal-poor island'' in the t-SNE 2D map \citep[see Figs 4 and 5 of][]{Hughes2022}.  Stellar parameters (T$_{\rm eff}$, log~$g$, [Fe/H]) were then estimated for these stars via fits of model synthetic spectra to the blue channel data in the vicinity of H$\beta$ and in the regions containing the Fe~{\sc i} lines noted above.  The outcome was a set of 54 nominally EMP candidates, i.e. stars for which the estimated abundance satisfies [Fe/H] $\leq$ --3.0; these stars occupy the ``upper edge'' of the ``metal-poor island'' in the t-SNE 2D map \citep[see Fig.\ 6 of][]{Hughes2022}.

Overall, there are three distinct advantages to the approach.  First, as outlined in \citet{Buder2021}, the GALAH sample is essentially unbiased, as stars are targeted for observation in the GALAH survey via simple selection criteria.  These are: all stars with 12.0 $<$ $V_{JK}$ $<$ 14.0 (where $V_{JK}$ is a proxy for the $V$ magnitude calculated from 2MASS photometry, see \citet{Buder2021}), declination $<$ +10$\degr$ and $|b|$ $>$ 10$\degr$ in regions of the sky that have at least 400 targets in $\pi$ square degrees (the field of view of the 2dF multi-fibre instrument).  Consequently, the GALAH EMP candidate list will, for example, include candidate EMP main sequence turnoff stars that are excluded in red giant-focussed EMP star searches, such as the SkyMapper EMP program where a colour-cut is used to select against stars warmer than T$_{\rm eff}$$\approx$ 5750 K \citep{DaCosta2019}.  The potential presence of EMP turnoff stars in the GALAH sample is in fact complimentary to surveys such as the TOPoS survey that specifically targets EMP turnoff stars \citep{Caffau2013}.  Detailed study of EMP main sequence stars can provide chemo-dynamical information that adds to that from EMP red giants.   Such stars are also important, for example, for the study of the complex behaviour of Li abundances at low [Fe/H] \citep[e.g.,][]{Bonifacio2018}.  

The second advantage of the nature of the GALAH survey target list is that it will include any carbon-enhanced metal-poor (CEMP) stars meeting the selection criteria.  CEMP-stars, especially CEMP-no stars that are not enhanced in $s$-process elements \citep[see][for CEMP-star type definitions]{Beers2005}, are increasingly common with decreasing [Fe/H] values, such that for [Fe/H] $\la$ --4.0, essentially all stars are C-rich \citep[see][and the references therein]{Norris2019}.  The inclusion, or exclusion, of CEMP-stars in photometry-based EMP searches such as the SkyMapper \citep{DaCosta2019} and Pristine \citep{Starkenburg2017}
programs is a complicated function of [C/Fe], [Fe/H] and effective temperature (and potentially also [O/Fe] and [N/Fe]) that is not easily quantified.  However, such selection uncertainties are not an issue with the simple apparent-magnitude selection criteria employed in GALAH. 

The third advantage of using the GALAH survey as a potential source of EMP candidates is that any confirmed new EMP stars will, with $V$ $\la$ 14, be relatively easy to study further in great detail. This is in contrast to the majority of EMP stars that are fainter, often significantly so, and hence require substantial resources to study in detail.  The relative brightness of any EMP stars in GALAH could allow, for example, the determination of oxygen and nitrogen abundances, which require high-resolution spectra in the near-uv ($\lambda$3100--3200\AA\/ for the OH-bands, and $\lambda$3360\AA\/ for the NH-bands), abundances for an increased number of neutron-capture elements, and studies of isotopic ratios such as $^{12}$C/$^{13}$C and those of the isotopes of magnesium.

In the following section the selection of the stars targeted for observation is outlined, followed by a description of the observations and analysis procedures.  The results are presented in \S 3 including a discussion of the selection efficiency of the t-SNE approach, and of the metallicity distribution function for the stars observed.  A comparison of the spectrophotometric temperatures and surface gravities with those from the GALAH DR3 database is given in \S\S3.2, while \S\S 3.3 gives information on a small number of noteworthy stars.  The final section summarizes the outcomes. 

\section{Sample Definition and Observations}

A set of statistically selected GALAH DR3 EMP candidates for potential low-resolution spectroscopic follow-up observation, 
containing 93
%confirmed Aug4
stars (after removing one duplicate entry), was provided by Arvind Hughes in August 2021. %(Hughes, 2021, priv.\ comm.).  
Using the facilities provided by the Simbad and Vizier databases\footnote{http://simbad.cds.unistra.fr/simbad/ \citet{Wenger2000}, https://vizier.cds.unistra.fr/viz-bin/VizieR \citet{Ochsenbein2000}}, 16 %confirmed Aug 4
 of the stars in the list were found to have existing metallicity estimates of sufficient precision (e.g., observed at high dispersion) to exclude them from the follow-up observing program.  The remaining 77 %confirmed Aug 4
candidates lack any existing metallicity information, which includes any [Fe/H] estimates in GALAH DR3, as these are unreliable for [Fe/H] $\la$ --2 because of the limited wavelength coverage of the GALAH spectra.  There are, however,  a small number of stars in the data set for which metallicity estimates are available from LAMOST DR5 \citep{Luo2019} and RAVE DR5 \citep{Kunder2017}.     All 77 stars in this set, which we will refer to as the ``2021 list'', %Of this set of stars, 75 %confirmed Aug 4
have been observed with the ANU 2.3m telescope (see below).  %these numbers are OK, no includes extra two, previously not observed.

For completeness, we note that the on-line version of Table 2 in \citet{Hughes2022} lists instead 53 distinct GALAH DR3 EMP candidates (the duplicate is repeated).  Four of the stars have high dispersion [Fe/H] estimates \citep[see Table 1 of ][]{Hughes2022}, while again using the facilities of the Simbad and Vizier databases, there are an additional three stars in the list possessing high dispersion metallicity determinations (see Table \ref{tab:tab1}).  In addition, star GALAH DR3 ID 180126003201229 is most likely a highly reddened OB-star, as it is included as such in the catalogues of \citet{MohrSmith2017} and \citet{Zari2021}.  The Gaia DR3 catalogue \citep{Gaiacollab2022} lists T$_{\rm eff}$ = 15270 K, log~$g$ = 3.51 for the star.  It was observed as part of the auxiliary programs in GALAH and, since it lies at $b$ = --0.085$\degr$, it is strictly not part of the GALAH DR3 sample. 
%lies at $b$ = --0.085$\degr$ so it is not clear why it is even in the GALAH DR3 catalogue.  
This star and those with existing high dispersion metallicity estimates have been excluded, leaving 45 stars, which we will refer to as the ``Table 2'' list of GALAH EMP candidates.  Of these 45 candidates, all but six are included in the ``2021 list''; however, all six of these stars have also been observed at the 2.3m telescope.
%with one of these six, star 160328004201370, having been observed at the 2.3m as part of the SkyMapper EMP search program
%\citep[][and on-going]{DaCosta2019}.   The parameters for this star from the spectrophotometric fit to the observed spectrum (see below) are 
%T$_{\rm eff}$ = 4500 K, log~$g$ = 1.375 and [Fe/H] = --3.125, noting that the spectrophotometric fit excluded the region in the vicinity of the G-band of CH, as the observed G-band was considerably weaker than in the model spectrum.  
We therefore have observations of all 77 candidates in the ``2021 list'' and of all 45 candidates in the ``Table 2 list'' for a total sample of 83 stars with 2.3m telescope observations.  % 45 -- 6 not in 2021 list -- 2 in 2021 list but not observed = 37 with spectra  plus 1 with 2.3m spectra = 38 with 2.3m spectra. These numbers are OK.
Given the large degree of overlap between the two samples we will mostly focus on results from the combined sample of 83 stars with 2.3m telescope observations.

\begin{table*}
\caption{Details of additional stars in Table 2 of \citet{Hughes2022} with published [Fe/H] values from high dispersion spectroscopy.}
\label{tab:tab1}
\begin{tabular}{lllcl}
\hline
Gaia DR3 Source   & GALAH DR3 ID & Other ID & [Fe/H] & Reference\\
\hline
%170805005101110 & HE 0048--6408 & --3.75 & \citet{Placco2014} \\
%170906004601108 & CS 22953-0003 & --2.93 & \citet{Yong2013} \\
%161118004701028 & SMSS J051008.62--372019.8 & --3.20 & \citet{Jacobson2015}\\ 
3824738543969555840 & 160421002101189 & HE 0926--0508 & --2.78 & \citet{Barklem2005} \\
6041027705498286464 & 160415004601352 & SMSS J155730.10--293922.6 & --2.77 & \citet{Yong2021} \\
%170904000601186 & CS 30322-023 & --3.39 & \citet{Masseron2010}\\
6403844758583266684 & 170711005101182 & SMSS J213402.81--622421.1 & --3.12 & \citet{Yong2021} \\
\hline
\end{tabular}
\end{table*}

The observations were carried out between August 2021 and August 2022 with the ANU 2.3m telescope at Siding Spring Observatory, using the WiFeS integral field spectrograph \citep{Dopita2010}.  The B3000 grating was employed with the blue camera to obtain resolution $R$ $\approx$ 3000 spectra that cover the wavelength range $\lambda3400-5800$\AA.  Exposure times were set to obtain a signal-to-noise ratio (S/N) of at least 20 per pixel at the H and K lines of Ca~{\sc ii}.  The integral field nature of the WiFeS instrument enabled useful spectra to be obtained even in poor seeing conditions.   In these respects the observations are completely consistent with those described in \citet{DaCosta2019} for the SkyMapper search for EMP stars.  Indeed some of the GALAH DR3 EMP candidates had already been observed in the SkyMapper EMP program.

The raw observed spectra were extracted, sky-subtracted and wavelength-calibrated, and then flux-calibrated via observations of a number of well-established flux standards \citep[see][for a more detailed discussion of the flux-calibration process]{DaCosta2019}.  
In particular, all stars are observed with the atmospheric dispersion direction parallel to the IFU slits to mininize flux loss, and the use of an IFU spectrograph minimizes any seeing losses.  Two or more observations of at least four flux standards are conducted each night.  We find that the (relative) absolute flux calibration is stable on timescales of nights to months.   In particular, the slope from $\lambda$4000 to 5800 \AA\/ in the fluxed spectra is very well defined.  This slope is sensitive to both effective temperature and gravity for an assumed reddening.

The {\sc fitter} code \citep[see][for a description]{Norris2013} was used to determine the best estimate of T$_{\rm eff}$, log~$g$ and [Fe/H] for each star via a comparison of the observed fluxes with those from a set of model atmospheres, given an (iterated if necessary) estimate of the star's reddening.  Details of the fitting procedure are described in \citet{DaCosta2019}.  The only significant change from \citet{DaCosta2019} is that instead of MARCS 1D model atmosphere fluxes \citep{Gustafsson2008}, we have used an extension of the grid of model fluxes discussed in \citet{Nordlander2019}.  Specifically, for giants (log~$g$ $\leq$ 3.5) the models employ $v_{mic}$  = 2 km~s$^{-1}$ and are computed in spherical symmetry, while for dwarfs $v_{mic}$  = 1 km~s$^{-1}$ and plane-parallel symmetry is used.  As noted in \citet{Nordlander2019}, atomic lines in the model spectra are a selection from VALD3 \citep{Ryabchikova2015}, with molecular lines primarily from \citet{Brooke2013, Brooke2014}, \citet{Masseron2014}, 
\citet{Ram2014} and \citet{Sneden2014}.  Examples of the spectrophotometric fits are shown in Fig.\ 5 of \citet{DaCosta2019}.
The temperatures, gravities and metallicities derived from the spectrophotometric fits will be designated T$_{\rm eff~2.3m}$, log~$g_{2.3m}$ and [Fe/H]$_{2.3m}$, respectively.

The T$_{\rm eff~2.3m}$, log~$g_{2.3m}$ and [Fe/H]$_{2.3m}$ values are quantized at the $\pm$25~K, $\pm$0.125 dex and $\pm$0.125 dex level, and the best-fit parameters are generally well determined since the stars are relatively bright.  %As noted in \citet{DaCosta2019}, 
Consequently, the uncertainty in the appropriate reddening value is frequently the largest contributor to the uncertainty in the derived temperature; errors arising from uncertainty in the flux calibration are generally less significant.  Overall, we estimate that typical uncertainties are $\pm$100 K, 0.3 -- 0.35 dex and 0.25 -- 0.3 dex in T$_{\rm eff~2.3m}$, log~$g_{2.3m}$ and [Fe/H]$_{2.3m}$, respectively.  Finally, we note that region in the vicinity of the G-band (CH) at $\lambda$ $\sim$ 4300\AA\/ was excluded from the fit if the observed band strength deviated substantially from the model spectrum, which is computed with [C/Fe] = 0 dex.  Since only 3 of the GALAH EMP candidates are found to have strong CH-features (see \S \ref{sect:3.3}), the [Fe/H]$_{2.3m}$ values for the bulk of the sample, given the 0.25 -- 0.3 dex uncertainty, are not affected by the assumption of a solar [C/Fe] ratio.  For the C-rich stars, it is possible that the large non-solar [C/Fe] values affect the structure of the stellar atmospheres compared to [C/Fe] = 0 models, but the size of any effect on [Fe/H]$_{2.3m}$ is not easily quantified.  It is possible that the [Fe/H]$_{2.3m}$ values for these 3 stars are uncertain by an additional 0.25 -- 0.3 dex.

Table \ref{tab:tab2} shows, for the first five stars, the Gaia DR3 Source identifier \citep{Gaiacollab2016,Gaiacollab2022}, the GALAH DR3 identifier \citep{Buder2021}, the Gaia DR3 J2000 right ascension (RA) and declination (Dec), the G-magnitude and B$_{\rm P}$--R$_{\rm P}$ colour (also from Gaia DR3) for each star together with the T$_{\rm eff~2.3m}$, log~$g_{2.3m}$, [Fe/H]$_{2.3m}$ and reddening values E(B--V)$_{2.3m}$ from the spectrophotometric fits to the observed spectra.  The full list of these parameters for all 83 stars with 2.3m observations is available in the online supplementary material.

\begin{table*}
	\centering
	\caption{Properties of the GALAH stars observed at the ANU 2.3m telescope.  Columns are Gaia DR3 source ID, GALAH DR3 identifier, RA and Dec (J2000), Gaia DR3 G mag and B$_{\rm P}$--R$_{\rm P}$ colour, and the effective temperature T$_{\rm eff~23m}$, surface gravity log~$g_{23m}$, and metallicity [Fe/H]$_{23m}$ from the spectrophotometric fits.  Complete table available in the online supplementary material.}
	\label{tab:tab2}
	\begin{tabular}{llcccccccc} 
\hline
Gaia DR3 Source   &    GALAH DR3 ID    &       RA     &     Dec     &      G   &  B$_{\rm P}$--R$_{\rm P}$ & T$_{\rm eff~2.3m}$ & 
log~$g_{2.3m}$ & [Fe/H]$_{2.3m}$ & E(B--V)$_{2.3m}$ \\ 
                  &                    &    (J2000)   &   (J2000)   &     mag  &     mag           &  K &  cgs  & dex & mag\\ 
\hline
4973234033240265088 & 161118002601383  &   00 06 06.68 & --51 32 40.6   & 12.874 & 0.937  &  5175  &    2.250  &  --2.250 & 0.01\\
4709807608617381248 & 170805005101087  &   00 53 07.82 & --63 45 01.5   & 13.421 & 0.896  &  5325  &    2.875  &  --2.250 & 0.02\\
4678520146255821952 & 150903002402109  &   04 14 16.25 & --60 09 04.8   & 12.857 & 1.054  &  4950  &    1.625  &  --2.500  & 0.04\\
4679059387989727872 & 150903002402348  &   04 19 20.01 & --58 40 42.4   & 12.321 & 0.954  &  5350  &    2.125  &  --2.750 & 0.02\\
4675029093759082496 & 161011003401196   &  04 22 53.12  & --65 22 31.4   & 13.575 & 1.077  &  4975  &    2.000  &  --2.750 & 0.02\\
\hline
	\end{tabular}
	\\{\it Note:} The T$_{\rm eff~2.3m}$, log~$g_{2.3m}$ and [Fe/H]$_{2.3m}$ values listed are the best-fit values from the spectrophotometric fitting process.  The values are quantized at the $\pm$25~K, $\pm$0.125 dex and $\pm$0.125 dex level but this does not indicate that the parameters are accurate to this level of precision.
\end{table*}

\section{Results}

\subsection{Selection efficiency and the Metallicity Distribution}

Overall the $t$-SNE statistical selection approach discussed in \citet{Hughes2022} has proved to be quite successful in identifying metal-poor stars.  Of the 83 candidates in the combined sample with spectrophotometric metallicity estimates from the 2.3m observations, only 6 have [Fe/H]$_{2.3m}$ \mbox{$>$ --2.0} for a small false positive or contamination rate of  7 $\pm$ 3 per cent.  This indicates an efficient selection process, indeed the contamination rate is similar to that for the SkyMapper search for EMP stars, which also has a contamination rate of 7\% \citep{DaCosta2019}.  Further, the contamination rate does not appear to depend on stellar type: all 17 main sequence turnoff star candidates selected have [Fe/H]$_{2.3m}$ \mbox{$\leq$ --2.0} dex.  Further, approximately one-third of the candidates have [Fe/H]$_{2.3m}$ \mbox{$\leq$ --2.75}, which is again broadly consistent with the SkyMapper search where the fraction of candidates with 
[Fe/H]$_{2.3m}$ \mbox{$\leq$ --2.75} is $\sim$40\% \citep{DaCosta2019}.  However, only five of the 83 stars observed have sufficiently low [Fe/H]$_{2.3m}$ values to be considered likely genuine EMP stars, i.e., [Fe/H]$_{2.3m}$ $\leq$ --3.0; four of these stars are red giants and one is a main sequence turnoff star.  The red giant star with the lowest abundance has [Fe/H]$_{2.3m}$ = --3.5; details of this star and for others of interest are given in \S \ref{sect:3.3}.

\begin{figure}
\includegraphics[width=\columnwidth]{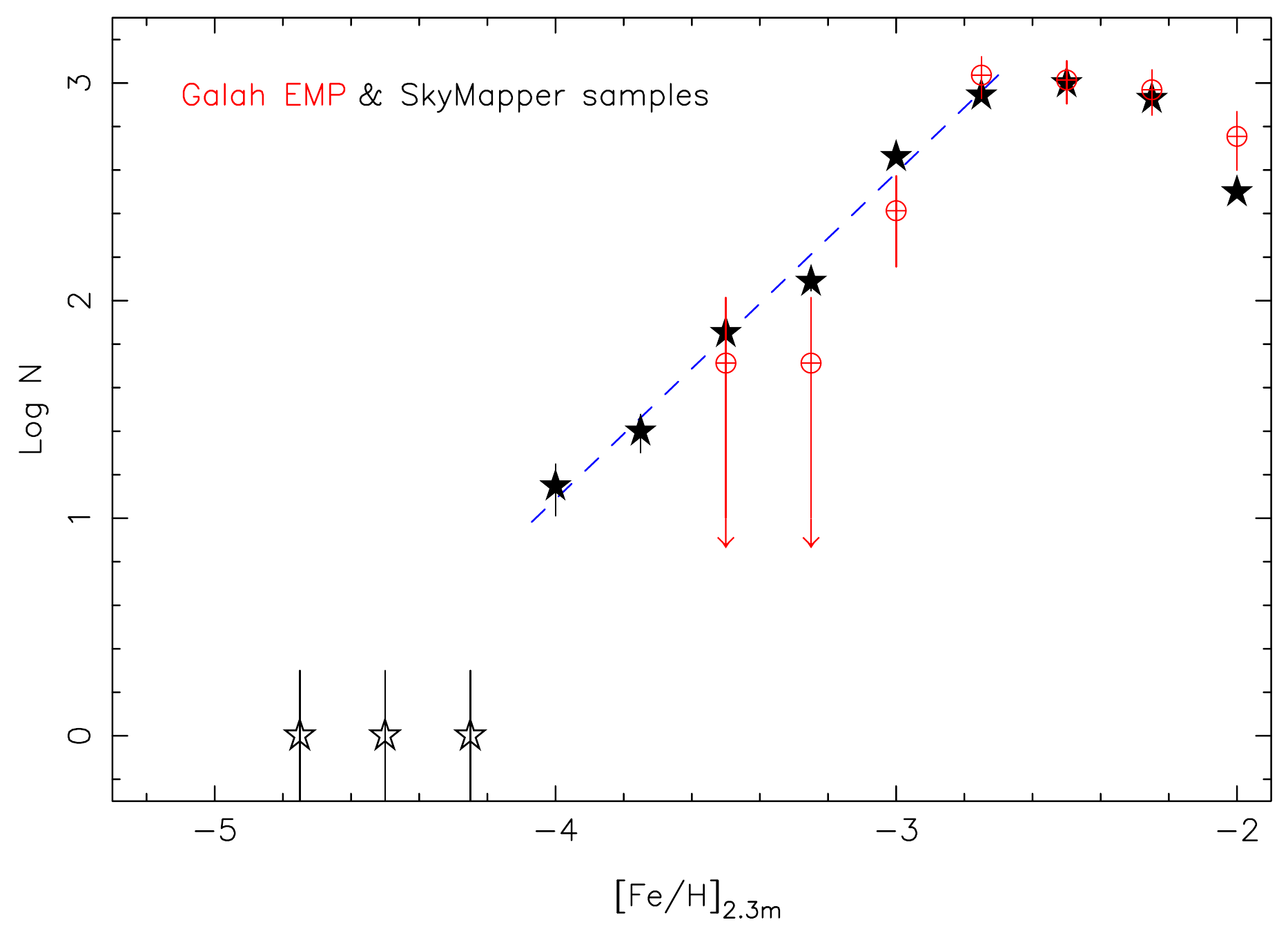}
    \caption{A comparison of the metallicity distribution function for all GALAH candidates with 2.3m spectrophotomtery (red circles) 
    with that for the SkyMapper
    sample \citep[][and unpublished]{DaCosta2019}, shown as black star symbols.   The GALAH numbers have been normalized to the SkyMapper numbers by using the number of stars with 
    [Fe/H]$_{2.3m}$ $\leq$ --2.25 dex.  Error bars are calculated assuming Poisson statistics for the actual star numbers, with the normalization applied to the GALAH $\pm$1$\sigma$ values.  Since the two most metal-poor metallicity bins for the GALAH candidates each contain only one star, the lower error bar strictly extends to --infinity, but has been truncated.  The blue dashed line has a slope of --1.5.  The three open star symbols are all C-rich objects.}
    \label{fig:MDF}
\end{figure}

The low return of genuine EMP candidates (5/83 or $\sim$6\%) seems at first a disappointing outcome, but in fact, given the small sample size and the extreme rarity of EMP stars, such an outcome is not unexpected.  We illustrate this as follows.  In Fig.\ \ref{fig:MDF} we show the metallicity distribution function (MDF) of all the GALAH candidates with 2.3m spectra, i.e., including the main sequence turnoff stars, and compare it to the MDF for the much larger sample of stars observed in the SkyMapper program \citep[][and unpublished]{DaCosta2019}.  In both cases the [Fe/H]$_{2.3m}$ values have been binned in increments of 0.25 dex, and we note that because of the colour selection employed \citep{DaCosta2019}, the SkyMapper sample is giant-focussed; main sequence turnoff stars are excluded.  The GALAH candidate distribution has been normalized to that for the SkyMapper sample by comparing the numbers of stars in the two samples with [Fe/H]$_{2.3m}$ $\leq$ --2.25 dex.  Despite the small number of stars in the current sample, it is evident that the MDF for the GALAH candidates is consistent with that for the SkyMapper sample and with the MDF slope of --1.5 for --4.0 $\leq$ [Fe/H]$_{2.3m}$ $\leq$ --2.75 \citep{DaCosta2019,Yong2021}.  

Given the normalization, the SkyMapper MDF predicts that the GALAH sample of 83 stars should have had 13 EMP (i.e., 
[Fe/H]$_{2.3m}$ $\leq$ --3) stars, whereas only 5 have been detected.  Assuming the standard deviation for the number of stars in a 
given metallicity range follows Poisson statistics (i.e., $\sigma$(n) = 
$\sqrt{n}$), this 5 stars observed versus 13 stars predicted difference is 
1.9$\times$ the combined $\sigma$ and so not significant. The normalization also predicts less than one candidate below [Fe/H]$_{2.3m}$ = --3.5, so the lack of any such star is not surprising.  Finally, with only 17 main sequence turnoff stars in the full candidate sample, it is not possible to draw any definite conclusions, but the MDF for these stars alone is consistent with that for the remainder of the sample. 

For completeness, we note that when considering only the 45 stars with 2.3m spectrophotometry that are in the ``Table 2'' list \citep{Hughes2022}, the selection efficiency is somewhat improved: only two stars have [Fe/H]$_{2.3m}$ \mbox{$>$ --2.0} for a 4 $\pm$ 3\% contamination rate (cf.\ 7 $\pm$ 3\% for the full set) while 19 have Fe/H]$_{2.3m}$ \mbox{$\leq$ --2.75}, or 42 $\pm$ 10\% (cf.\ 31 $\pm$ 6\%).  All 5 probable genuine EMP stars are in the ``Table 2'' list for a return of 11 $\pm$ 5\% versus 6 $\pm$ 3\% for the full set of observed stars.  While the differences are relatively minor, the set of candidates in Table 2 of \citet{Hughes2022} is an improvement over the candidate list originally supplied.  We should also not lose sight of the fact that the t-SNE statistical approach did re-identify $\sim$20 very and extremely metal-poor stars that were previously known, in addition to those used to identify the ``metal-poor island'' in the t-SNE 2D map.  {\it The statistical approach is therefore very efficient at selecting EMP candidates}.

\subsection{Comparison with GALAH DR3 parameters}

\begin{figure}
\includegraphics[width=\columnwidth]{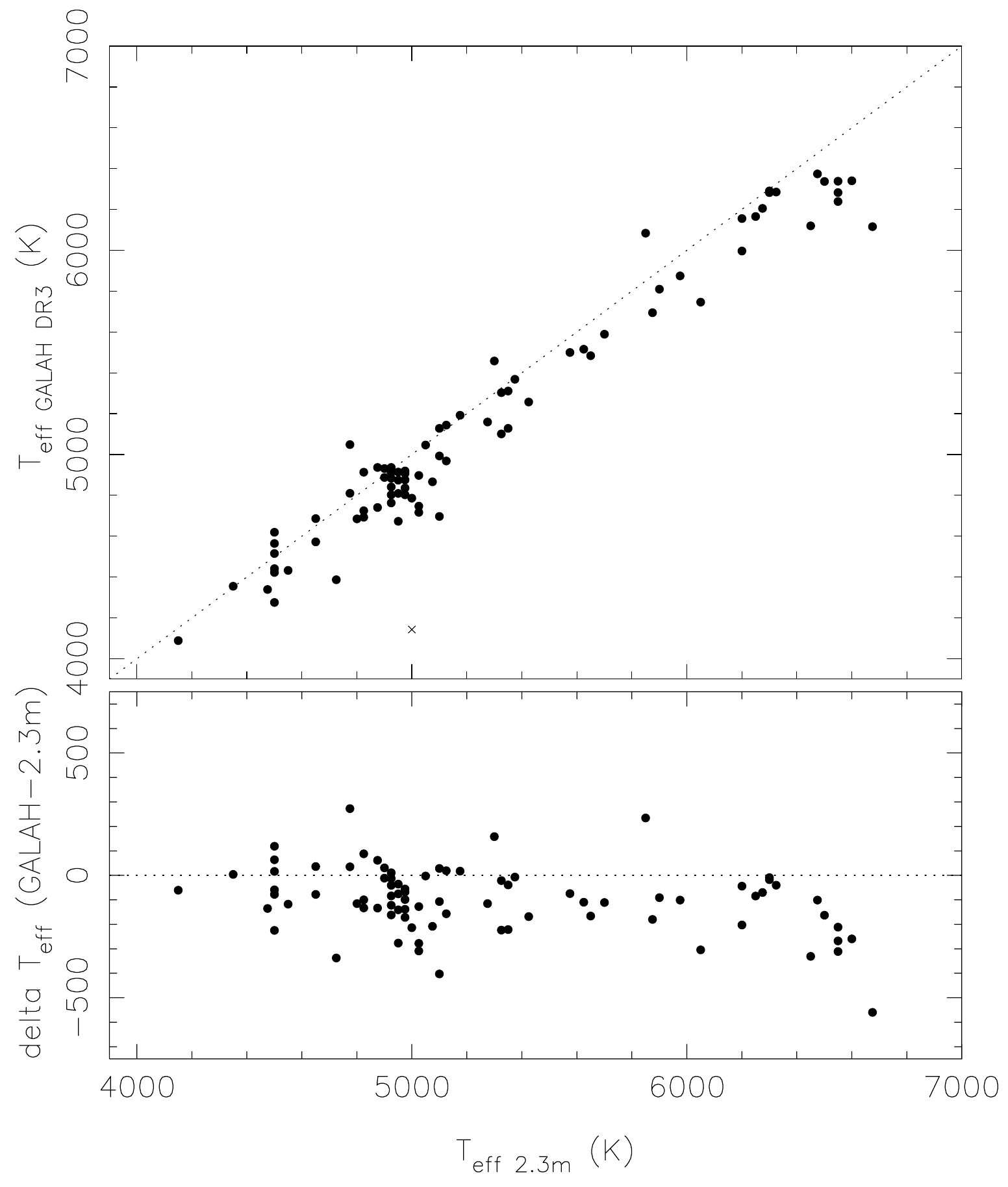}
    \caption{{\it Upper panel:} A direct comparison of the effective temperatures derived from the spectrophotometric fits to the observed 2.3m spectra,  T$_{\rm eff~2.3m}$, with those, T$_{\rm eff~GALAH~DR3}$, tabulated in the GALAH DR3 release.  The dotted line is the 1:1 line and the x-symbol is for the star where the GALAH DR3 temperature is substantially cooler than the spectrophotometric estimate.  {\it Lower panel:} The difference between the GALAH DR3 values and those derived from the spectrophotometry.  The dotted line is for zero difference.}
    \label{fig:Temp_comp}
\end{figure}

In Fig.\ \ref{fig:Temp_comp} we show a comparison of the effective temperatures derived from the spectrophotometric fits with the temperatures listed in the GALAH DR3 database for all the stars with 2.3m observations.  One star, GALAH DR3 ID 140116004301131,
has a listed GALAH DR3 T$_{\rm eff}$ value that is clearly significantly different (860~K cooler) from the value (5000~K) derived from the 2.3m spectrophotometry\footnote{We note that a T$_{\rm eff}$ value of order 5000 K is consistent with the colours of this star, while the lower GALAH DR3 T$_{\rm eff}$ value is not. For example, $(g-i)_0$ = 0.67 from SkyMapper DR2 \citep{Onken2019} corresponding to T$_{\rm eff}$ $\approx$ 5050 K for a metal-poor red giant. Given the unusual size of the temperature difference, the star may well be worthy of more detailed follow-up.}.  For the remaining 82 stars the mean difference in T$_{\rm eff}$, in the sense (GALAH DR3 -- 2.3m) is --101 K with a standard deviation of 135 K\@.  For these predominantly metal-poor stars, the GALAH DR3  T$_{\rm eff}$ values are set primarily by 1D model atmosphere fits to the H$\alpha$ and H$\beta$ line profiles \citep[see][]{Buder2021}, while the spectrophotometric temperatures are primarily set by the Balmer continuum slope \citep[after correction for reddening; see the discussion in][]{DaCosta2019}.  The consistency of the two temperature scales is then reassuring, modulo the small systematic offset.  We note, however, that in Fig.\ \ref{fig:Temp_comp}, beyond T$_{\rm eff~2.3m}$ $\approx$ 6200~K, the GALAH DR3 T$_{\rm eff}$ values for the sample tend to be systematically underestimated relative to the spectrophometric temperatures.  It is unclear why this is the case though it may be related to the continuum normalization of the GALAH spectra (Buder, 2022, priv.\ comm.).

As regards uncertainties, although the 2.3m spectrophotometric T$_{\rm eff~2.3m}$ values are quantized at the 25 K level, in practice uncertainty in the adopted reddening and in the fitting process means that the actual uncertainties in the spectrophotometric T$_{\rm eff~2.3m}$ values are of order 100~K\@.  If that's the case, then the standard deviation of the differences would imply that the GALAH DR3 temperatures for these metal-poor stars also have uncertainties of order 100 K\@.  Reassuringly, excluding the star with GALAH DR3 ID 160530003301334 for which there is no entry in the GALAH DR3 database for the error in T$_{\rm eff}$ (or for the error in log~$g$), 
as well as star 140116004301131, the mean GALAH DR3 T$_{\rm eff}$ error for the remaining 81 stars is 104 K with a standard deviation of 22 K\@.  

\begin{figure}
\includegraphics[width=\columnwidth]{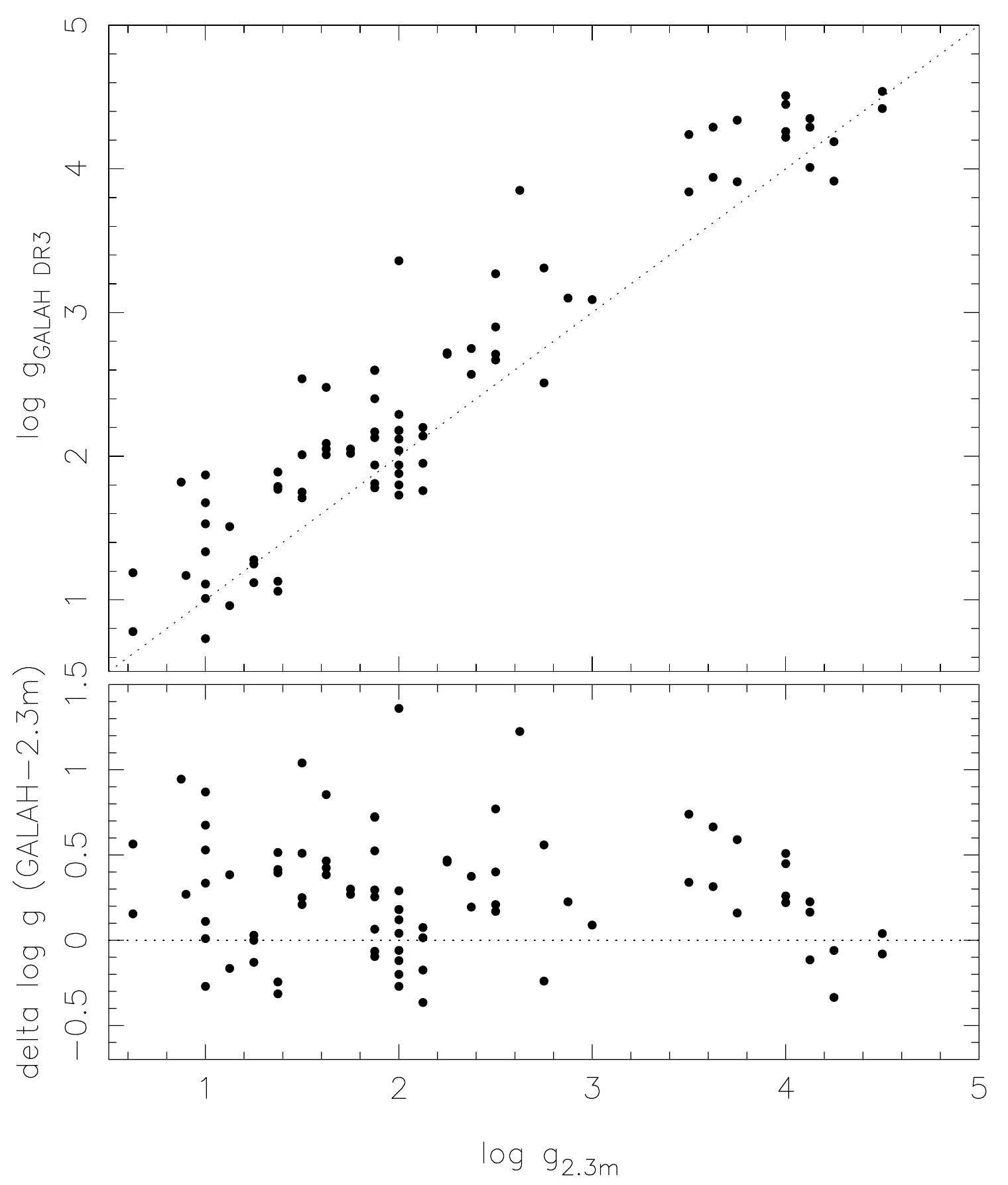}
    \caption{{\it Upper panel:} A direct comparison of the surface gravities derived from the spectrophotometric fits to the observed 2.3m spectra, log~$g_{2.3m}$, with those, log~$g_{\rm GALAH~DR3}$, tabulated in the GALAH DR3 release.  The dotted line is the 1:1 line.  {\it Lower panel:} The difference between the GALAH DR3 values and those derived from the spectrophotometry.  The dotted line is for zero difference.}
    \label{fig:logg_comp}
\end{figure}

In a similar fashion, we show in Fig.\ \ref{fig:logg_comp} a comparison of the surface gravities (log~$g_{2.3m}$) derived from the 2.3m spectrophotometry with those given in the GALAH DR3 database (log~$g_{GALAH DR3}$).  The log~$g_{2.3m}$ values are primarily set, for a given T$_{\rm eff~2.3m}$ and assumed reddening, by the model spectral fit to the Balmer jump in the observed spectrum
\citep[see the discussion in][]{DaCosta2019}.  No assumption regarding the distance to the star is required.  On the other hand, the GALAH DR3 surface gravities are estimated \citep[see][]{Buder2021} by applying bolometric corrections, appropriate for the adopted temperatures, to the reddening corrected 2MASS $K$ magnitudes.  Stellar masses then are estimated from appropriate isochrones, and crucially, luminosities are calculated by adopting the distances estimated from Gaia DR2 parallax measurements as described in \citet{BailerJones2018}.  Given the relatively bright apparent magnitude cutoff for the GALAH sample ($V$ $\approx$ 14), this approach generally works well for the dwarfs that dominate the GALAH sample \citep[see][]{Buder2021} as these stars are relatively nearby with well-determined parallaxes.  However, this approach could result in significant uncertainty in the derived log~$g_{GALAH DR3}$ values for luminous red giants that are likely to be at significant distances and whose parallaxes may well be uncertain \citep[see, e.g., the discussion in][]{Cordoni2021}.

Fig.\ \ref{fig:logg_comp} shows that the two surface gravity estimates are, in fact, well correlated but with the GALAH DR3 values  systematically higher than the log~$g_{2.3m}$ estimates.  The mean difference in log~$g$, in the sense (GALAH DR3 -- 2.3m) is 0.27 with a standard deviation of 0.36 dex for the 83 stars in the full sample.  As for the T$_{\rm eff}$ values, the origin of the systematic offset in the log~$g$ values is unclear.  The spectrophotometric log~$g_{2.3m}$ values are quantized at the 0.125 dex level but a realistic estimate of the uncertainty in the values is 0.3 -- 0.35 dex.  This value comes from \citep{DaCosta2019} and is the dispersion in  log~$g_{2.3m}$ at fixed T$_{\rm eff~2.3m}$ about a metal-poor RGB isochrone for the SkyMapper EMP star sample.  In principle therefore, the uncertainties in the log~$g_{2.3m}$ values could be responsible for the entire scatter in the log~$g$ differences.  However, the same could be said for the GALAH DR3 values as the average error in the log~$g$ values listed in the GALAH DR3 database for these stars is 0.30 with a standard deviation of 0.12 dex.  There is, nevertheless, some indication that the likely larger distances and therefore larger uncertainties in the GALAH DR3 log~$g$ values for the red giants do play a role.  For the 17 stars in Fig.\ \ref{fig:logg_comp} with log~$g$ exceeding 3.0, i.e., subgiants and main sequence turnoff stars, the standard deviation of the differences is 0.30 dex, whereas for the 66 stars with log~$g$ $\leq$ 3, i.e., the giants, the standard deviation of the differences is significantly larger at 0.37 dex.  The mean offset, however, remains essentially unchanged (0.24 vs 0.28).

\subsection{Stars of interest} \label{sect:3.3}

\begin{figure}
\includegraphics[width=\columnwidth]{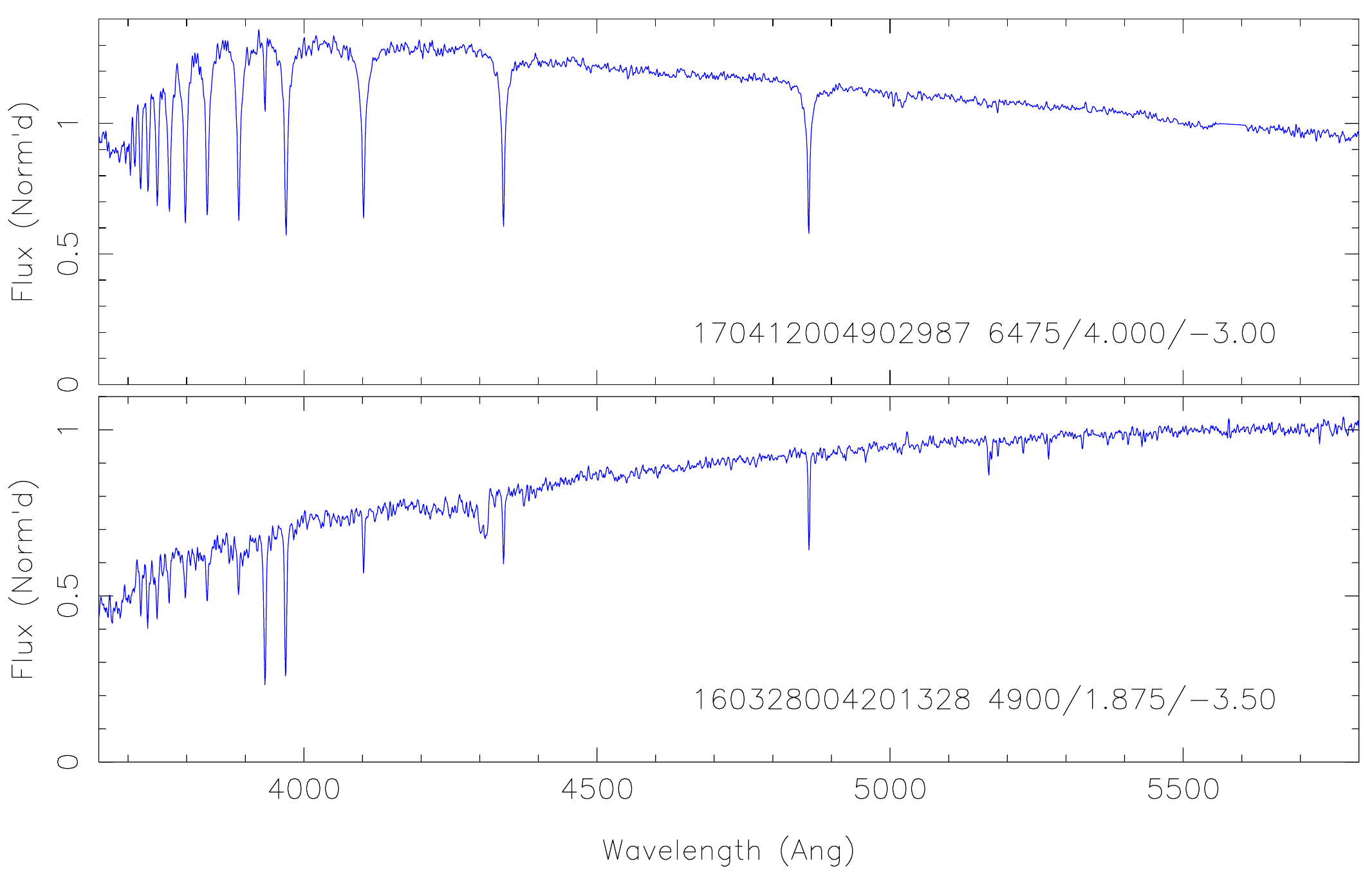}
    \caption{Examples of EMP stars confirmed in this study.  The upper panel shows the 2.3m spectrum of the main sequence turnoff star GALAH DR3 ID 170412004902087 with flux normalized to unity at 5500\AA.    The lower panel shows the 2.3m spectrum of the red giant star GALAH DR3 ID 160328004201328 with flux again normalized to unity at 5500\AA.  The T$_{\rm eff~2.3m}$, log~$g_{2.3m}$ and [Fe/H]$_{2.3m}$ values derived from the spectrophometric fits to the spectra are given in the lower right of each panel.}
    \label{fig:empstars}
\end{figure}

In Fig.\ \ref{fig:empstars} we show the 2.3m spectra for two of the five candidates found to be likely genuine EMP stars, 
i.e., [Fe/H]$_{2.3m}$ $\leq$ --3.0.  Star GALAH DR3 ID 170412004902987 is evidently an EMP main sequence turnoff star with T$_{\rm eff~2.3m}$ = 6475 K, log~$g_{2.3m}$ = 4.0 and [Fe/H]$_{2.3m}$ = --3.00.  For comparison, the GALAH DR3 database entry for this star gives consistent values, namely T$_{\rm eff}$ = 6374 K and log~$g$ = 4.26.  The second star depicted in Fig.\ \ref{fig:empstars} is the most metal-poor object in the set of 2.3m observations.  This star is GALAH DR3 ID 160328004201328 and it is evidently an EMP red giant with the spectrophotometric fitting process yielding T$_{\rm eff~2.3m}$ = 4900 K, log~$g_{2.3m}$ = 1.875 and [Fe/H]$_{2.3m}$ = --3.50.  Again we emphasize that values given are those of the spectrophotometric fit and that the listed precision does not imply comparable accuracy in the parameters. The GALAH DR3 database entry for this star gives T$_{\rm eff}$ = 4931 K and log~$g$ = 0.50, values again consistent with the 2.3m determinations.  We note that the other three probable EMP stars in the observed are red giants: these 
are GALAH DR3 ID 160514003301102 with T$_{\rm eff~2.3m}$ = 4500 K, log~$g_{2.3m}$ = 1.00 and [Fe/H]$_{2.3m}$ = --3.00, 160328004201370 with T$_{\rm eff2.3m}$ = 4500 K, log~$g_{2.3m}$ = 1.375 and [Fe/H]$_{2.3m}$ = --3.125 and 140312003501132 with T$_{\rm eff2.3m}$ = 4725 K, log~$g_{2.3m}$ = 0.625 and [Fe/H]$_{2.3m}$ = --3.25.  The GALAH DR3 database entries for these stars are again consistent with the 2.3m values, namely: T$_{\rm eff}$ = 4275 K, log~$g$ = 1.11, T$_{\rm eff}$ = 4517 K, log~$g$ = 1.13, and T$_{\rm eff}$ = 4387 K, log~$g$ = 1.19, respectively.

\begin{figure}
\includegraphics[width=\columnwidth]{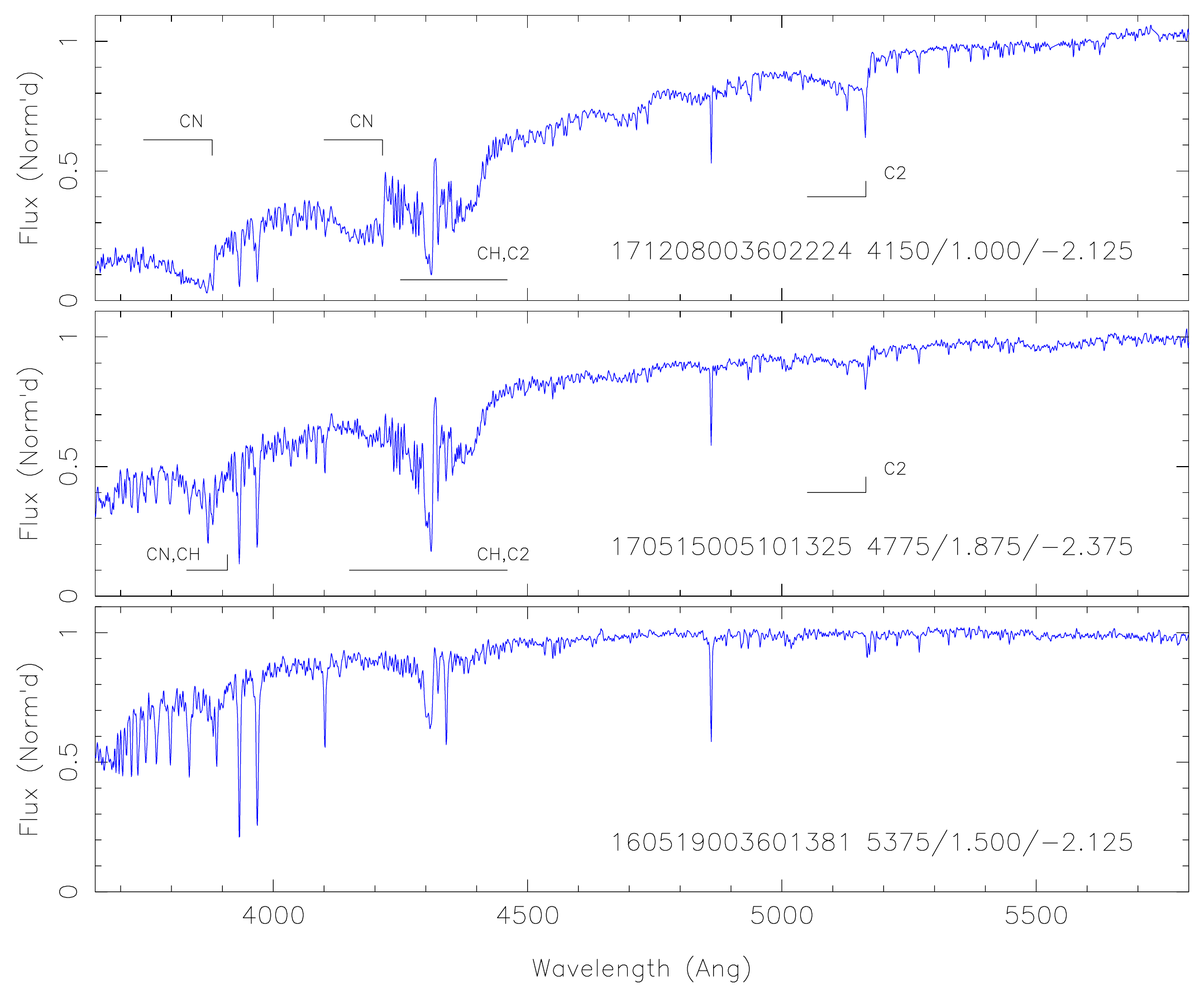}
    \caption{{\it Upper panel:} The 2.3m spectrum of the star GALAH DR3 ID 171208003602224 with flux normalized to unity at 5500\AA. Strong features of CN, CH and C$_2$ are marked. {\it Middle panel:} The 2.3m spectrum of the star GALAH DR3 ID 170515005101325 with flux normalized to unity at 5500\AA.  Strong features of CN, CH and C$_2$ are again marked; note that the CN-bands at 3883\AA\/ and 4215\AA\/ are noticeably weaker in this star than in star 171208003602224.   {\it Lower panel:} The 2.3m spectrum of the star GALAH DR3 ID 160519003601381 with flux normalized to unity at 5500\AA. The T$_{\rm eff~2.3m}$, log~$g_{2.3m}$ and [Fe/H]$_{2.3m}$ values derived from the spectrophometric fits to the spectra are given in the lower right of each panel.  }
    \label{fig:cstars}
\end{figure}

There are three other stars in the sample worthy of note.  These are GALAH DR3 IDs 171208002602224, 170515005101325 and 160519003601381 whose 2.3m spectra are shown in the panels of Fig.\ \ref{fig:cstars}.  All of these stars, which are in Table 2 of \citet{Hughes2022}, are evidently significantly enhanced in carbon.  Initial spectrum synthesis calculations, employing the same approach as outlined in \S 3.4 of \citet{DaCosta2019}, reveal that for both 171208002602224 and 170515005101325, [C/Fe] $>$ +1, while for 160519003601381 [C/Fe] $\approx$ +0.8 dex.  The synthetic spectra \citep[see][for details]{Nordlander2019} were calculated using a 1D, LTE approximation with the atmospheric parameters derived from the 2.3m spectra\footnote{The GALAH DR3 parameters for these stars are T$_{\rm eff}$ = 4089, 5048 and 5375 K, and log~$g$ = 1.33, 2.17 and 1.50, respectively.}, and were convolved to match the resolution of the 2.3m spectra.  Using the definition of \citet{Aoki2007}, and noting that in each case [Fe/H] $<$ --2.0, the stars qualify as CEMP stars.  

Star 171208002602224 is HE~0520--5012 and is a known carbon star, as it is in the \citet{Christlieb2001} catalogue of faint high latitude carbon stars.  However, it lacks any previous metallicity estimate prior to the value [Fe/H] $\approx$ --2.125 derived here.  Given the strength of the CN-bands at $\lambda$3883 and $\lambda$4215\AA\/ in the 2.3m spectrum, it is likely that this star is strongly enhanced in nitrogen as well as in carbon.  %Preliminary spectral synthesis calculations suggest [N/Fe] $\sim$1.5 dex.  

\begin{figure}
\includegraphics[width=\columnwidth]{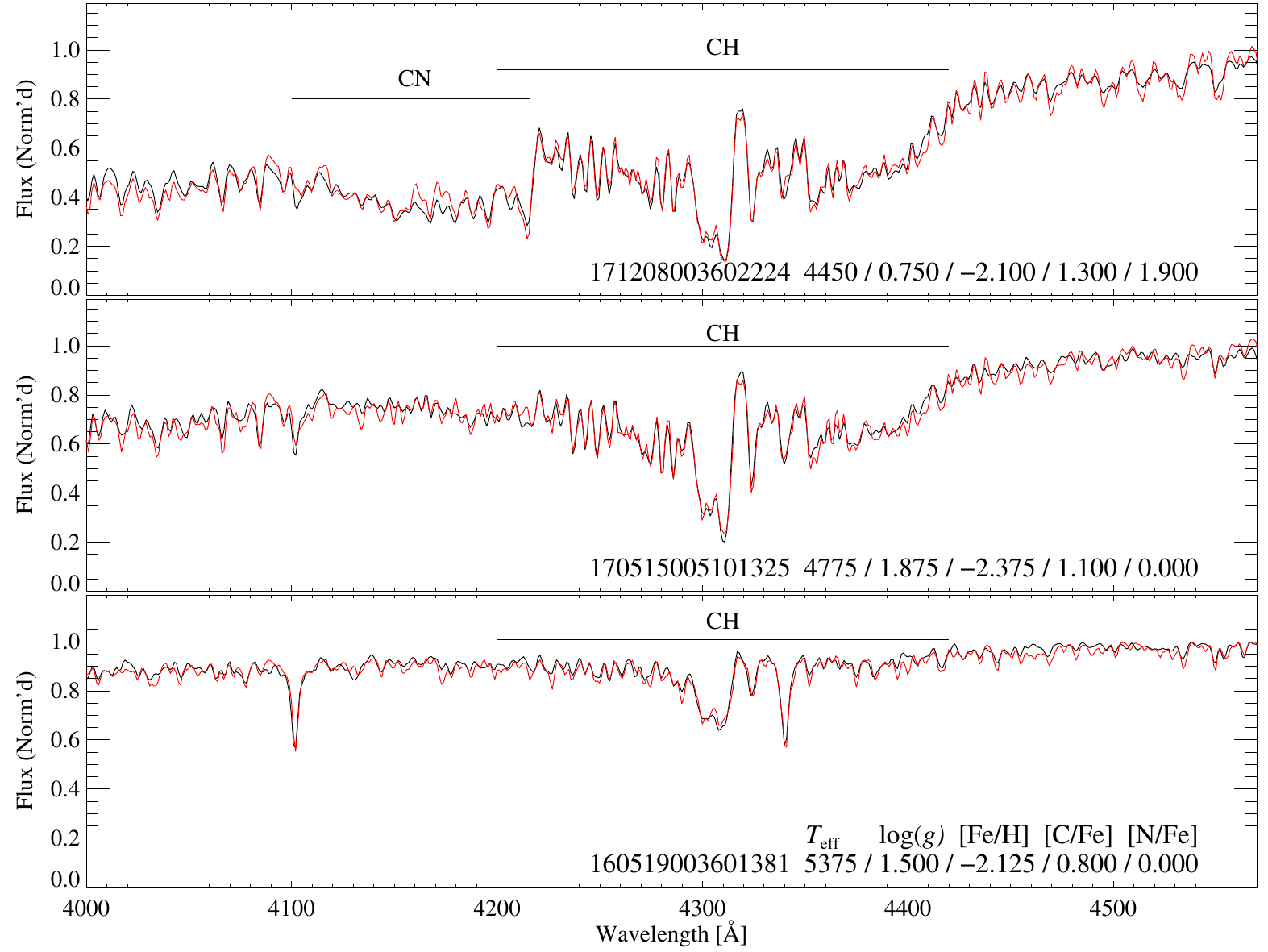}
 \caption{Comparison of the observed (black) and synthetic (red) spectra for GALAH DR3 stars 171208003602224 (top), 170515005101325 (middle) and 160519003601381 (bottom). For 171208003602224 the derived values for [C/Fe] and [N/Fe] are +1.3 and +1.9, respectively, and are uncertain at the $\pm$0.3 level. The location of the CN and CH features are indicated by the horizontal bars. For the middle and bottom two stars the [C/Fe] values are +1.1 and +0.8, respectively, with an uncertainty at the $\pm$0.2 level. Here, $\rm [N/Fe] = 0$ is assumed rather than determined. In each panel, following the DR3 ID, the adopted $T_{\rm eff}$, $\log g$, [Fe/H], [C/Fe] and [N/Fe] values are listed.}
   \label{fig:cfits}
\end{figure}

We show in Fig.\ \ref{fig:cfits} the outcomes of detailed synthetic spectrum fits in which the best-fit value of [C/Fe], and of [N/Fe] for 171208002602224, has been determined by $\chi^{2}$-minimization for the comparison between the observed and the synthetic spectra.  For 160519003601381 and 170515005101325 the model atmosphere parameters were those from the 2.3m spectrophotometry, and the synthetic spectra were calculated with [N/Fe] = 0.  The resulting [C/Fe] values are +0.8 $\pm$ 0.2 and +1.1 $\pm$ 0.2, respectively, confirming the initial estimates.  The case for 171208002602224 is more complex because of the large abundances of C and N\@.  The fit shown for this star employs T$_{\rm eff}$ = 4450~K, some $\sim$300~K hotter than the 2.3m spectrophotometric determination.  Similarly, the log $g$ value is 0.25 lower at 0.75, while the [Fe/H] value is unchanged. Given the probable complex nature of the atmosphere of this star, these differences in atmospheric parameters are likely within the uncertainties in the parameters.   The best-fit shown is for [C/Fe] = +1.3 and [N/Fe] = +1.9; both values are likely uncertain at the $\pm$0.3 level.

Stars 170515005101325 and 160519003601381 are likely the first CEMP-$s$ stars identified in the GALAH DR3 dataset as both stars show enhanced abundances for $s$-process elements in the GALAH DR3 database.  Specifically, for 170515005101325, [Y/Fe] = 0.38 $\pm$ 0.13, [Ba/Fe] = 1.16 $\pm$ 0.09, [Nd/Fe] = 1.46 $\pm$ 0.14 but there is no measurement for [Eu/Fe].  For 160519003601381, [Y/Fe] = 0.26 $\pm$ 0.19, [Ba/Fe] = 1.35 $\pm$ 0.10, [La/Fe] = 1.59 $\pm$ 0.18 and there is only a weak upper limit for [Eu/Fe] ([Eu/Fe] $<$ 1.8).  It is likely that both stars are CEMP-$s$ stars but confirmation of the classification requires a measurement of [Eu/Fe], as the CEMP-$s$ classification requires [Ba/Fe] $>$ 1.0 and [Ba/Eu] $>$ +0.5 dex \citep[e.g.,][]{Beers2005}.  For star 171208002602224 no information is given in the GALAH DR3 database for any neutron-capture elements.  The occurrence of these C-rich stars in the sample of EMP candidates with 2.3m observations confirms that the underlying t-SNE selection process is not affected by the C-rich (or not) nature of the stars, which is not surprising since there are no strong carbon features in the wavelengths covered by the GALAH spectra.

The high [C/Fe] and [$s$/Fe] abundances in CEMP-$s$ stars are generally thought to have their origin in mass-transfer from an asymptotic giant branch (AGB) star in a binary system \citep[e.g.,][]{Ryan2005}, where the AGB star is now a white dwarf.  Judging by the [Ba/Fe] versus [Fe/H] panel in Fig.\ 21 of \citet{Buder2021}, and requiring [Ba/Fe] $\ga$ +1 and [Fe/H] $\la$ --2.0, it appears likely that there are a small number of additional CEMP-$s$ candidates in the GALAH DR3 sample.  Such a sample could be readily followed-up with low-resolution spectroscopy to confirm the CEMP nature (although [Eu/Fe] measurements would still be required for a definite classification).  

We also note that the one of the ``Value-Added Catalogues'' to the GALAH DR3 release contains the kinematics of the GALAH stars \citep[see][for details]{Buder2021}.  The majority of the stars considered here have (when the kinematic information is available) kinematics consistent with membership of the ``inner halo'', i.e., z$_{\rm max}$ (maximum height above the plane) greater than $\sim$2 kpc and R$_{apo}$ $\la$ 12 kpc.  Stars 160514003301102 ([Fe/H]$_{2.3m}$ = --3.0) and 170515005101325 (CEMP-$s$ candidate) have retrograde orbits.

\section{Conclusions}

In this study have we followed up with low resolution spectroscopy a number of candidate extremely metal-poor stars 
(i.e., [Fe/H] $\leq$ $\mbox{--3.0)}$ identified in the GALAH DR3 database.  The observed spectra provide values of T$_{\rm eff~2.3m}$, log~$g_{2.3m}$ and [Fe/H]$_{2.3m}$ for the candidates, which were identified from among the $\sim$600,00 stars with GALAH DR3 spectra through the application of the t-SNE statistical method \citep{Hughes2022}.  The statistical approach highlights candidate metal-poor stars, which are revealed as making up less than 1\% of the total sample, and the most promising EMP candidates are then identified via model synthetic spectral fitting to narrow wavelength intervals in the HERMES blue channel spectra \citep{Hughes2022}.  The approach has been successful both in (re-)identifying known metal-poor stars and in flagging candidates for confirmation via low resolution spectroscopy.  

Observations of the list of previously unstudied EMP candidates has shown low contamination from metal-richer stars, and a metallicity distribution function consistent with previous work.  Five previously unknown probable  [Fe/H] $\leq$ --3.0 stars are revealed, four of which are red giants while one is a main sequence turnoff star.  The outcome of the statistical approach is clearly limited only by the extreme rarity of EMP stars in surveys, such as GALAH, that are not specifically aimed at finding low metallicity stars.  The simple selection criteria of the GALAH survey (and the lack of strong carbon features in the HERMES spectral bands) nevertheless means that EMP candidates selected are not biased by effective temperature or by carbon abundance, biases that can affect other EMP search programs.  Indeed the low resolution spectroscopy has revealed the first two probable CEMP-$s$ stars in the GALAH sample, as well as re-identifying a known high-latitude carbon star.  The relatively bright ($V$ $\la$ 14) nature of the GALAH stars also means detailed follow-up with full wavelength coverage high dispersion spectrographs would be straightforward.

In essence the results here demonstrate that the statistical approach adopted in \citet{Hughes2022} should be equally productive when applied to forthcoming even larger stellar spectroscopic surveys that will reach fainter magnitudes, such as those proposed with the WEAVE \citep{Dalton2012,Dalton2020}, 4MOST \citep{deJong2019} and {\bf DESI \citep{Cooper2022}} instruments.

\section*{Acknowledgements}

This research has been supported in part by the Australian Research Council Centre of Excellence for All Sky Astrophysics in 3 Dimensions (ASTRO 3D) through project number CE170100013. It has also made use of the SIMBAD database,
operated at CDS, Strasbourg, France, and the VizieR catalogue access tool, CDS, Strasbourg, France (DOI: 10.26093/cds/vizier). The original description of the VizieR service was published in 2000, A\&AS 143, 23.

This work has made use of data from the European Space Agency (ESA) mission
{\it Gaia} (\url{https://www.cosmos.esa.int/gaia}), processed by the {\it Gaia}
Data Processing and Analysis Consortium (DPAC,
\url{https://www.cosmos.esa.int/web/gaia/dpac/consortium}). Funding for the DPAC
has been provided by national institutions, in particular the institutions
participating in the {\it Gaia} Multilateral Agreement.

We acknowledge the traditional owners of the land on which the ANU 2.3m telescope is located, the Gamilaraay people, and pay our respects to elders past, present and emerging.

%%%%%%%%%%%%%%%%%%%%%%%%%%%%%%%%%%%%%%%%%%%%%%%%%%
\section*{Data Availability}

The underlying data will be shared on reasonable request to the authors.

%%%%%%%%%%%%%%%%%%%% REFERENCES %%%%%%%%%%%%%%%%%%

% The best way to enter references is to use BibTeX:

\bibliographystyle{mnras}
\bibliography{Galah} % if your bibtex file is called example.bib

% Alternatively you could enter them by hand, like this:
% This method is tedious and prone to error if you have lots of references
%\begin{thebibliography}{99}
%\bibitem[\protect\citeauthoryear{Author}{2012}]{Author2012}
%Author A.~N., 2013, Journal of Improbable Astronomy, 1, 1
%\bibitem[\protect\citeauthoryear{Others}{2013}]{Others2013}
%Others S., 2012, Journal of Interesting Stuff, 17, 198
%\end{thebibliography}

%%%%%%%%%%%%%%%%%%%%%%%%%%%%%%%%%%%%%%%%%%%%%%%%%%

%%%%%%%%%%%%%%%%% APPENDICES %%%%%%%%%%%%%%%%%%%%%

%%%%%%%%%%%%%%%%%%%%%%%%%%%%%%%%%%%%%%%%%%%%%%%%%%

% Don't change these lines
\bsp	% typesetting comment
\label{lastpage}
\end{document}